# A discrete model for the growth and spread of the Scottish populations of red squirrels (*Sciurus vulgaris*) and grey squirrels (*Sciurus carolinensis*).

Jean-Baptiste Gramain, Institute of Mathematics, University of Aberdeen, King's College, Fraser Noble Building, Aberdeen AB24 3UE, UK

Corresponding author: Jean-Baptiste Gramain, jbgramain@abdn.ac.uk

**Abstract:**

In this article, a model, discrete in space and time, is developed to describe the growth and spread of the Scottish populations of red squirrels (*Sciurus vulgaris*) and grey squirrel (*Sciurus carolinensis*). The initial state for the model is designed using a large dataset of records of sightings of individuals of both species reported by members of the public. Choices of parameters involved in the model and their values are informed by the analysis of this dataset for the period 2011-2016, and model predictions are compared to records for the years 2006-2019.

## 1. Introduction:

The North American grey squirrel (*Sciurus carolinensis*) is non-native to the UK, and was introduced in various parts of Britain in the early 1900's. It has since spread throughout England, Wales and Southern Scotland, apparently outcompeting and displacing (or even locally driving to extinction) the indigenous red squirrel (*Sciurus vulgaris*). Several models have been designed to describe the changes in red and grey squirrel population sizes and distributions (see e.g. Okubo *et al.*, 1989; Rushton *et al.*, 1997; Tompkins *et al.*, 2003), but the reasons behind the replacement of red squirrels by grey squirrels remain unclear (Reynolds, 1985; Tompkins *et al.*, 2003).

Three main factors have been studied in order to understand the population dynamics of these two species: interspecific competition, habitat preference and environmental change, and disease. While there is some evidence that, for example in areas dominated by deciduous trees, grey squirrels have the ability to exclude red squirrels (see e.g. Gurnell *et al.*, 2004), it has been shown that both species can coexist for long periods in coniferous areas (Bryce *et al.*, 2002). Furthermore, analysis of available data for Norfolk between 1960 and 1982 (Reynolds, 1985) as well as further model simulations (Okubo *et al.*, 1989; Rushton *et al.*, 1997) indicate that interspecific competition alone cannot explain the rate of displacement of red squirrels by grey squirrels. Reynolds (1985) also found that habitat change was unlikely to explain the decline of red squirrels in Norfolk. The role of disease, in particular squirrel parapoxvirus, also remains unclear. There is evidence suggesting that squirrel parapoxvirus (SQPV) was introduced in the UK with the arrival of grey squirrels (Scott *et al.*, 1981; Duff *et al.*, 1996; Sainsbury and Ward, 1996). Grey squirrels are thought to act as a reservoir for SQPV (Sainsbury *et al.*, 2000), and the virus doesn't appear to have any strong impact on their health while it is known to cause severe disease and death in red squirrels (Tompkins *et al.*, 2002). SQPV contamination of red squirrels by grey squirrels has thus been studied as a contributive factor in the decline of red squirrel populations, but conclusions on the strength and scale of this process vary (Reynolds, 1985; Tompkins *et al.*, 2003; Bruemmer *et al.*, 2010; Chantrey *et al.*, 2014). It should also be noted that, to date, SQPV doesn't appear to be present in Scotland (Sainsbury *et al.*, 2000), and will therefore not be included in the present study.

In response to the decline in UK red squirrel populations, various monitoring and conservation programmes have been put in place, such as Saving Scotland's Red Squirrels (https://scottishsquirrels.org.uk/) which collates reports of squirrel sightings, and uses traps to catch grey squirrels (which are then tagged or culled). Other programmes in Wales and Scotland aim to use the native pine marten (*Martes martes*) to control the grey squirrel populations, as red squirrels have a greater ability to avoid predation by pine martens (Twining *et al.*, 2021; Roberts and Heesterbeek, 2021; Slade *et al.*, 2023).

Given the difficulty to study in detail the various interactions between the two squirrel species (and their pathogens and predators), whether experimentally or through observations, modelling provides a useful and practical way to explore population dynamics and derive predictions. This often takes one of two main forms (Rushton *et al.*, 1997): attempts to relate species' distributions to environmental factors for which geographical data is available (such as tree cover density or forest type), or the use of models for population dynamics (such as Lotka-Volterra models for interspecific competition or predator-prey systems, or SEIR models) in order to exhibit distribution patterns. In this study, a new model is described for the growth and spread of red and grey squirrel populations in Scotland, which aims to unify these two approaches. The model is based on a classical Lotka-Volterra model for interspecific competition, but is coupled with a model for dispersal based on habitat preference, including altitude, tree cover density and forest type. The model and its various parameters are inspired by and tested against a very large dataset of squirrel sighting reports available from the National Biodiversity Network.

## 2. Material and methods:

The model constructed to describe the evolution over time of the two Scottish squirrel populations is discrete in both time and space, and consists of two steps: interspecific competition and dispersal. Both steps will be described discretely, but would obviously occur continuously in the natural environment. An area covering Scotland was divided into ca. 10 000 *cells*. Within each cell, the two populations are subject to interspecific competition, and a portion of each population disperses each year to the neighbouring cells according to their *attractiveness* (defined in section 2.2.).

## 2.1. Area and grid.

The geographic area used for the model is delimited by the latitudes 53.955°N and 59.04°N, and by the longitudes -0.08°W and 7.04°W (see Fig. 1). This area completely contains Scotland, apart from part of the Outer Hebrides, where both squirrel species are absent. The region was divided into 10 057 cells $S_{i,j}$ ($0 \leq i \leq 112, 0 \leq j \leq 88$), the North-West corner of $S_{i,j}$ having latitude $54 + 0.045i$°N and longitude $0.08j$°W. In this way, the area shown in Fig. 1 is 113-cell long from South to North, and 89-cell long from East to West. Each cell $S_{i,j}$ has South-North side 0.045° of latitude and East-West side 0.08° of longitude, both of which are approximately 5km, and which give an area $a_{i,j}$ ranging from 23.05km² to 26.30km². The model describes the evolution over time of the numbers $R_t^{i,j}$ and $G_t^{i,j}$ of red and grey squirrels living in cell $S_{i,j}$ in year $t$.

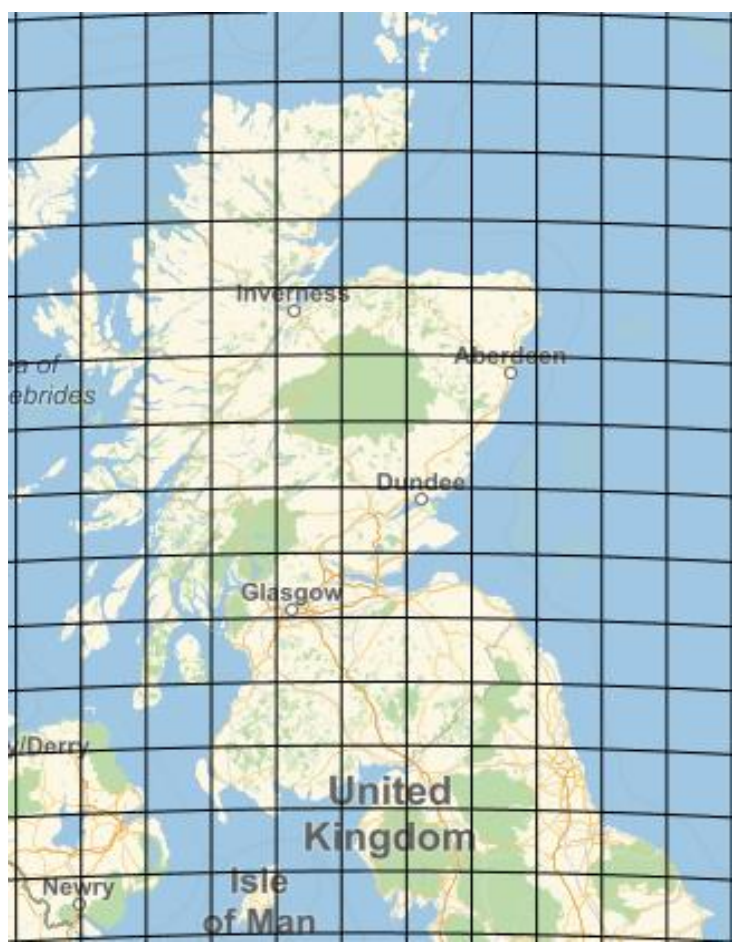

**Figure 2.** Map of the area considered for the model. A subset of the grid (defined by latitudes and longitudes) used in the model has been added.

## 2.2. Attractiveness.

To each $S_{i,j}$ are associated two numbers, $0 \leq A_R^{i,j}, A_G^{i,j} \leq 1$, which describe the *attractiveness* of $S_{i,j}$ to red and grey squirrels respectively. The terminology used should not be thought of as having any biological meaning, but rather a mathematical one. The attractiveness of a cell aims to capture the likelihood of individuals of each species to be present in or to disperse to the cell. For each cell and species, the attractiveness is in turn a function (defined later on) of the altitude $x_{i,j}$ of the cell $S_{i,j}$ (defined as the average of the altitudes of the four corners of $S_{i,j}$, computed using Wolfram Mathematica (Version 13.1.0.0)) and the forest type index $FTY_{i,j}$ of $S_{i,j}$.

## 2.3. Interspecific competition.

Interspecific competition between years $t$ and $t + 1$ is described by a discrete version of a classical Lotka-Volterra model as follows. For any cell $S_{k,l}$ with population sizes $R_t^{k,l}$ and $G_t^{k,l}$ in year $t$, let

$$IR_{t+1}^{k,l} = (1 + b_R - d_R)R_t^{k,l} \frac{(K_R^{k,l} - R_t^{k,l} - c_G G_t^{k,l})}{K_R^{k,l}},$$

$$IG_{t+1}^{k,l} = (1 + b_G - d_G)G_t^{k,l} \frac{(K_G^{k,l} - G_t^{k,l} - c_R R_t^{k,l})}{K_G^{k,l}},$$

where, for red squirrels, $b_R$ (respectively $d_R$) is the *per capita*, *per annum* birth rate (respectively death rate), $K_R^{k,l}$ is the carrying capacity of $S_{k,l}$ and $c_R$ is the competitive effect on grey squirrels, and similarly for grey squirrels, $b_G$, $d_G$, $K_G^{k,l}$ and $c_G$. The carrying capacities $K_R^{k,l}$ and $K_G^{k,l}$ are given by $\frac{a_{k,l}}{25} K_R$ and $\frac{a_{k,l}}{25} K_G$ respectively, where $K_R$ and $K_G$ are the carrying capacities for a 25km² area. Estimates for all these parameters are given in Table 1.

**Table 1.** Estimated values for the parameters used in the model. Birth rates and death rates are given *per capita per annum*, and carrying capacities are given for a 25 km² area. (Adapted from Tompkins *et al.* (2003))

| Parameter | Symbol | Value | Reference |
|---|---|---|---|
| *Both species* | | | |
| Death rate | $d_R = d_G$ | 0.4 | Gurnell (1987), Wauters *et al.* (2000) |
| Dispersal fraction | $\mu$ | 0.2 | Okubo *et al.* (1989) |
| | | | |
| *Red squirrel* | | | |
| Birth rate | $b_R$ | 1.0 | Okubo *et al.* (1989) |
| Carrying capacity | $K_R$ | 60 | Rushton *et al.* (1997) |
| Competitive effect on grey | $c_R$ | 0.61 | Bryce *et al.* (2001) |
| | | | |
| *Grey squirrel* | | | |
| Birth rate | $b_G$ | 1.2 | Okubo *et al.* (1989) |
| Carrying capacity | $K_G$ | 80 | Rushton *et al.* (1997) |
| Competitive effect on red | $c_G$ | 1.65 | Bryce *et al.* (2001) |

## 2.4. Dispersal.

For each $S_{k,l}$, the set $\mathcal{N}_{k,l}$ is defined to be the set of ordered pairs $(i,j)$ such that the cells $S_{i,j}$ and $S_{k,l}$ share exactly one or two corners. In this way, the elements of $\mathcal{N}_{k,l}$ label the *neighbours* of $S_{k,l}$, and, apart from those on the border of the region considered, each cell has 8 neighbours. After the interspecific competition step described above, a proportion $\mu$ (see Table 1) of each population disperses from $S_{k,l}$ to its neighbours, according to the *relative attractiveness* of each neighbour. For any $(i,j)$ in $\mathcal{N}_{k,l}$, the attractiveness to red and grey squirrels of $S_{i,j}$ relative to $S_{k,l}$ are given by

$$A_{R,(k,l)}^{i,j} = A_R^{i,j}/(\sum_{(m,n)\in\mathcal{N}_{k,l}} A_R^{m,n}) \text{ and } A_{G,(k,l)}^{i,j} = A_G^{i,j}/(\sum_{(m,n)\in\mathcal{N}_{k,l}} A_G^{m,n})$$

if these are defined, and

$$A_{R,(k,l)}^{i,j} = A_{G,(k,l)}^{i,j} = 0$$

otherwise. In particular, it is always the case that

$$\sum_{(i,j)\in\mathcal{N}_{k,l}} A_{R,(k,l)}^{i,j} = \sum_{(i,j)\in\mathcal{N}_{k,l}} A_{R,(k,l)}^{i,j} = 1,$$

unless all neighbours of $S_{k,l}$ have attractiveness 0. Also, if all neighbours of $S_{k,l}$ have equal attractiveness (for either squirrel species), then they also have equal attractiveness relative to $S_{k,l}$.

The numbers of red and grey squirrels which disperse from $S_{k,l}$ to any of its neighbours $S_{i,j}$ between years $t$ and $t+1$ are then given by

$$D_{R,(k,l)}^{i,j} = \mu A_{R,(k,l)}^{i,j} IR_{t+1}^{k,l} \text{ and } D_{G,(k,l)}^{i,j} = \mu A_{G,(k,l)}^{i,j} IG_{t+1}^{k,l}.$$

Note in particular that, unless all neighbours of $S_{k,l}$ have attractiveness 0,

$$\sum_{(i,j)\in\mathcal{N}_{k,l}} D_{R,(k,l)}^{i,j} = \mu IR_{t+1}^{k,l} \text{ and } \sum_{(i,j)\in\mathcal{N}_{k,l}} D_{G,(k,l)}^{i,j} = \mu IG_{t+1}^{k,l},$$

so that a proportion $\mu$ of each population does disperse from $S_{k,l}$ to its neighbourhood.

## 2.5. Overall population changes.

Putting together the two steps described above (interspecific competition and dispersal), the model predicts that, for any cell $S_{k,l}$, at the end of year $t$, the numbers of squirrels having left $S_{k,l}$ are $\mu IR_{t+1}^{k,l}$ and $\mu IG_{t+1}^{k,l}$, while the numbers of squirrels which moved into $S_{k,l}$ from its neighbouring cells are

$$D_R^{k,l} = \sum_{(i,j)\in\mathcal{N}_{k,l}} D_{R,(i,j)}^{k,l} \text{ and } D_G^{k,l} = \sum_{(i,j)\in\mathcal{N}_{k,l}} D_{G,(i,j)}^{k,l}.$$

This yields the following estimates for the populations of red and grey squirrels in $S_{k,l}$ in year $t+1$:

$$R_{t+1}^{k,l} = (1-\mu)IR_{t+1}^{k,l} + D_R^{k,l} \text{ and } G_{t+1}^{k,l} = (1-\mu)IG_{t+1}^{k,l} + D_G^{k,l}.$$

In order to implement this model, all that is left is to define the initial conditions $R_0^{i,j}$ and $G_0^{i,j}$ and the attractiveness $A_R^{i,j}$ and $A_G^{i,j}$ for each cell $S_{i,j}$ in the grid. This is done using existing data on the occurrence of red and grey squirrels in Scotland, as well as data on the type and extent of the woodland cover in the area studied.

## 2.6. Distribution data on red and grey squirrels

The National Biodiversity Network (NBN) shares data collected by various partners, such as the Scottish Wildlife Trust, the Centre for Environmental Data and Recording, the Biological Records Centre and the Mammal Society. For red and grey squirrels, this data consists of reports, made by organisations or members of the public, of squirrel sightings in the UK. This includes the date of the sighting, the number of individuals observed, and, when possible, the GPS coordinates of the sighting.

The use of such data as a proxy for animal abundance can lead to bias and errors (see e.g. MacKenzie *et al.*, 2002, 2009; McDonald-Madden *et al.*, 2010). For example, recording effort is likely to be inconsistent both in time and space, and biased towards red squirrels (as people tend to report "the rare" rather than "the common"). Data collected after March 2020 is also likely to diverge from data collected previously, as more people spent time outdoors following the emergence of COVID-19. Future studies could be conducted to examine such possible divergence, and more generally assess the relevance of such recording efforts to estimate trends in population abundances.

Data sets for red and grey squirrels recorded in the UK were obtained from the NBN Atlas (https://nbnatlas.org), and cleaned to suit the needs of this work.

Any unconfirmed record, any record without a precise date or location, and any record of a sighting made after 2019 was discarded. This yielded data sets of 95 652 red squirrel records and 704 476 grey squirrel records. When reported, abundance was ignored, and each record was only considered as evidence for the presence of one individual at the given time and location. For each record, the altitude of the record was computed and added to the data set using Wolfram Mathematica.

## 2.7. Initial and boundary conditions

For any cell $S_{i,j}$ located on the edge of the grid (i.e. $i \in \{0,112\}$ or $j \in \{0,88\}$), and for any cell $S_{i,j}$ with altitude $x_{i,j} < 0$, the initial numbers of red and grey squirrels were set to be $R_0^{i,j} = 0$ and $G_0^{i,j} = 0$ respectively. For any *inner cell*, the data sets described above were used to estimate the population sizes $R_0^{i,j}$ and $G_0^{i,j}$ as follows. The year 2006 was chosen to derive the initial condition for two reasons: it was the first year with a significant number of reports for both species in the region considered (3 227 and 2 915 for red and grey squirrels respectively), and it allowed the comparison between the model's predictions and actual data over several years. For each inner cell $S_{i,j}$ and each species, the number of 2006 records per km² was computed, and these "densities" were scaled linearly, to range from 0 to 1. The initial population $R_0^{i,j}$ (resp. $G_0^{i,j}$) was then defined as the product of the density by the carrying capacity $K_R^{i,j}$ (resp. $K_G^{i,j}$), reflecting the assumption that a cell with maximal density should be at carrying capacity for the corresponding species. The resulting estimates for the 2006 population sizes are represented in Fig. 2.

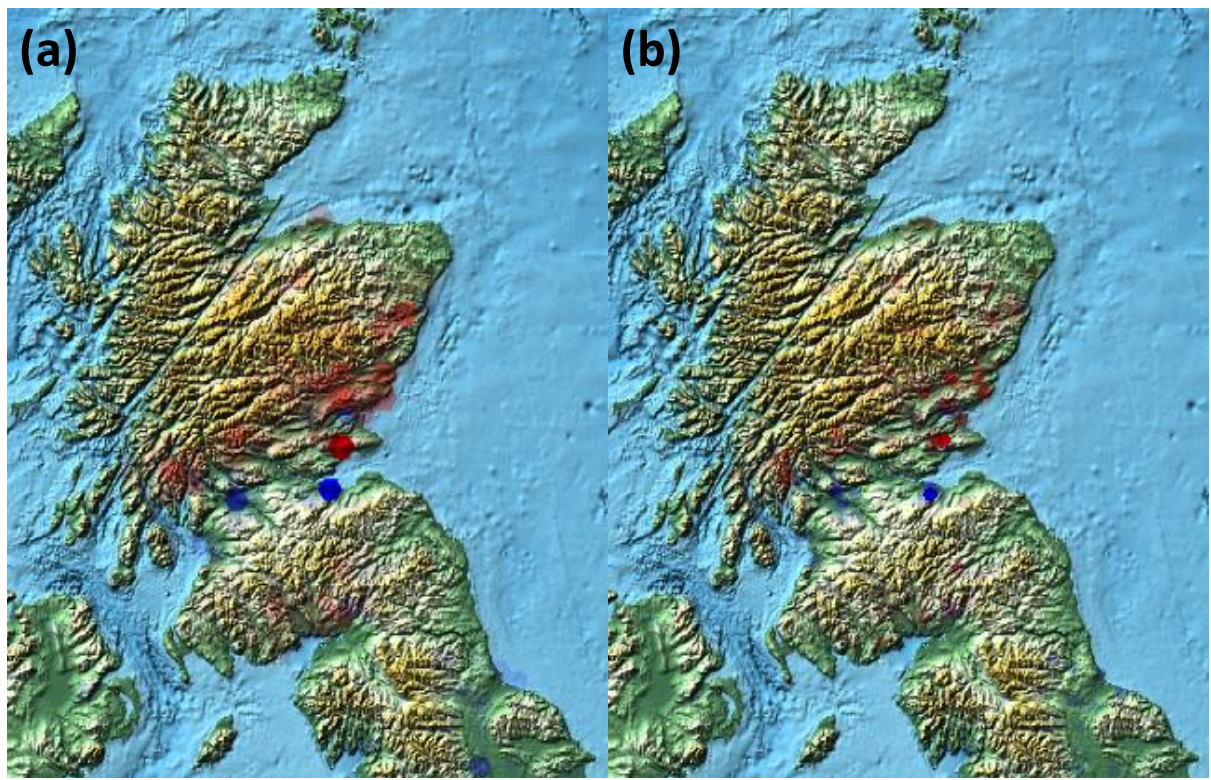

**Figure 2.** Initial conditions for the model. Red squirrels are represented in red, grey squirrels in blue. Colour opacity is proportional to numbers. (a) Number of records in Scotland in 2006. (b) Estimated population sizes in Scotland in 2006.

*2.8.   Woodland cover data*

In order to include habitat characteristics in the model, High Resolution Layers (HRL) from the Copernicus Land Monitoring Service, available from the European Environment Agency (EEA), were used. A total of eight images was downloaded and exploited, corresponding to the areas E30N40 and E30N30 (which, together, cover the whole of the UK), and representing the tree cover density and the forest type index for the years 2012 and 2015. For each year and each region, the images consist of pixels, each corresponding to a 100 m by 100 m square, and colour-coded to contain the data about the parameter considered (tree cover density or forest type index). Using Wolfram Mathematica, the tree cover density (TCD) for each pixel was recovered, expressed as the percentage of the corresponding square covered by trees. The forest type index (FTY) for each pixel was also recovered and expressed as one of four numbers: 1 for squares dominated by deciduous forests, 2 for squares dominated by coniferous forests, 1.5 for squares dominated by mixed forests, and 0 for squares with unknown dominating forest type.

Each image was produced by the EEA using the Projected Coordinate Reference System ETRS89 / ETRS-LAEA / EPSG: 3035 (with projection center's coordinates 52°N, 10°E, false easting 4 321 000 m and false northing 3 210 000 m). Using coordinate conversion formulas from the International Association of Oil & Gas Producers (IAOGP, 2019), the latitude and longitude of each pixel of the Copernicus High Resolution Layers was computed using Wolfram Mathematica.

For each cell $S_{i,j}$ in the overall grid, the (mean) tree cover density $TCD_{i,j}$ of $S_{i,j}$ was calculated (as a percentage) as the average of the TCDs for all the pixels lying in $S_{i,j}$. The forest type index $FTY_{i,j}$ of $S_{i,j}$ was also calculated as follows. For cells without tree cover, or where all pixels had FTY 0, the FTY of the cell was taken to be 0. For cells with at least one pixel with non-zero FTY, the FTY of the cell was defined as the weighted average of the non-zero FTYs of the pixels, the weights being given by the TCD of the corresponding pixels. In this way, each grid cell was given an FTY of 0, or between 1 and 2 (an FTY closer to 1 reflecting a tree cover dominated by deciduous trees, while an FTY closer to 2 reflected the dominance of coniferous trees). Such a weighted average was used in order to take into account the possibly different tree cover densities of the pixels considered when estimating the overall forest type of the cell. This was done separately for the years 2012 and 2015.

For each year between 2011 and 2016, the numbers of records of red and grey squirrels in each cell $S_{i,j}$ were also computed and recorded.

## 3. Results:

### 3.1. Attractiveness.

Analysis of the data described in sections 2.6. and 2.8. guided the definition, for each cell $S_{i,j}$ in the grid, of the attractiveness (to both red and grey squirrels) of $S_{i,j}$. For each cell and species, the attractiveness is a product of two factors, representing differences observed in the distribution of red and grey squirrels according to altitude and tree cover. The attractiveness of $S_{i,j}$ to red and grey squirrels are defined as

$$A_R^{i,j} = A_{R,Alt}^{i,j} A_{R,Tree}^{i,j} \text{ and } A_G^{i,j} = A_{G,Alt}^{i,j} A_{G,Tree}^{i,j}.$$

First, for any cell $S_{i,j}$ located on the edge of the grid or with altitude $x_{i,j} < 0$, the attractiveness to red and grey squirrels were set to be $A_R^{i,j} = A_G^{i,j} = 0$. Together with the initial conditions chosen for these cells, and given the description of dispersal, this ensures that, as the model runs, no red or grey squirrel will ever be present in any of these cells. This allows the model to keep within the boundaries of the region considered, and prevents the modelled spread of either squirrel population in areas actually covered by sea (and thus with negative altitude).

### 3.2. Altitude.

To define the components $A_{R,Alt}^{i,j}$ and $A_{G,Alt}^{i,j}$ of attractiveness corresponding to altitude for inner squares, the records of both species made in 2006 over the whole of the UK (3 572 records for red squirrels and 24 472 for grey squirrels) were used. For each species, the number of reports for each 100 ft altitude interval from 0 ft to 2 000 ft was computed, and these numbers rescaled linearly to range from 0 to 1. A slight modification of a Rayleigh distribution was then fitted by trial and error to define the attractiveness as a function of altitude (see Fig. 3). This was done to reflect the assumption that the abundance of squirrels of both species at a given altitude is positively correlated to the attractiveness of this altitude, i.e. that squirrels of both species are initially distributed according to attractiveness. The formulae obtained in this way are $A^{i,j}_{R,Alt} = f_r(x_{i,j})$ and $A^{i,j}_{G,Alt} = f_g(x_{i,j})$, where $x_{i,j}$ is the altitude of $S_{i,j}$, given in feet,

$$f_r(x) = \left(\frac{x+500}{500}\right) e^{\left(1-\left(\frac{x+500}{500}\right)^2\right)/2} \text{ and } f_g(x) = \left(\frac{x+250}{350}\right) e^{\left(1-\left(\frac{x+250}{350}\right)^2\right)/2}.$$

It should be noted (see Fig. 3) that, while $f_r$ doesn't give a particularly good fit at low altitudes, its tail differs significantly from that of $f_r$, reflecting the fact that a larger proportion of red squirrels than grey squirrels were found at higher altitudes. Indeed, 12.09% of the sightings of red squirrels occurred at an altitude of 700 ft or more, compared to only 2.62% for grey squirrels. At an altitude of 1 000 ft or more, these proportions decreased to 5.10% and 0.68% for red and grey squirrels respectively.

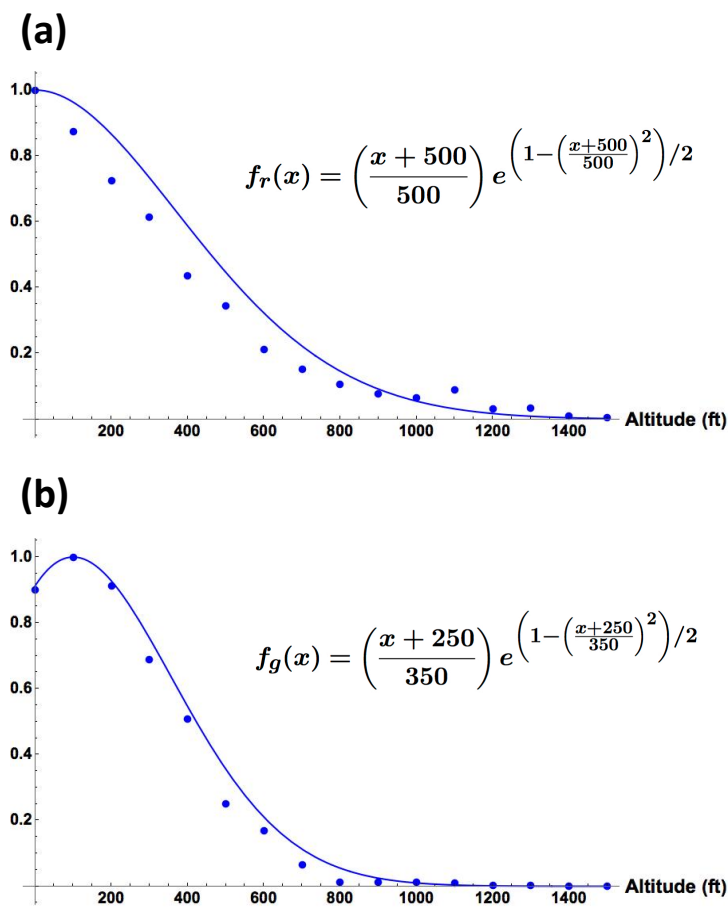

**Figure 3.** Plots of the altitude attractiveness functions (blue curves) defined to approximate the scaled distributions of 2006 records (blue dots). (a) Red squirrels. (b) Grey squirrels.

*3.3. Woodland cover.*

The tree cover densities and forest type indices computed for the 10 057 grid cells are summarised in Table 2. For both years (2012 and 2015), the minimum TCD for grid cells was 0%, and the minimum and maximum non-zero FTY were 1 and 2 respectively.

**Table 2.** The maximum tree cover density (TCD), mean TCD and mean non-zero forest type index (FTY) computed for the 10 057 grid cells of the region considered, using the Copernicus woodland cover data for the years 2012 and 2015.

| Data year | Maximum TCD (%) | Mean TCD (± SD) (%) | Mean non-zero FTY (± SD) |
|---|---|---|---|
| **2012** | 66 | 4.02 (± 8.27) | 1.54 (± 0.35) |
| **2015** | 67.8 | 4.20 (± 8.32) | 1.51 (± 0.34) |

The tree cover density and forest type index were found to be positively correlated, with the quadratic general linear model $TCD = 36.95 - 54.62 FTY + 24.60 FTY^2$ significantly explaining 30.18% of the variability ($p < 0.001$).

To describe the influence of the forest type index of a given cell $S_{i,j}$ on the probability of squirrels of each species being recorded in that cell, and thus define the components $A_{R,Tree}^{i,j}$ and $A_{G,Tree}^{i,j}$ of the attractiveness of $S_{i,j}$, the following analysis was carried out. Starting with the 2012 woodland cover data, cells with TCD 0% or FTY 0 were ignored, as were cells where no squirrel was recorded in the years 2011, 2012 and 2013. Similarly, cells with TCD 0% or FTY 0 using the 2015 woodland cover data were ignored, as were cells where no squirrel was recorded in the years 2014, 2015 and 2016. For the 2012 data, 8,317 cells were thus excluded (82.7%), while 8,103 cells were excluded for the 2015 data (80.6%). Note that, in this process, some cells may have been ignored using both the 2012 and 2015 woodland cover data, while some contributed twice to the remaining 3,694 cells, with possibly different FTYs for 2012 and 2015. Of the remaining 3,694 cells, 919 (24.9%) had records of squirrels of both species for these years and were also excluded. To estimate the probability of only red squirrels being recorded in a cell with a given FTY, the remaining 2,775 cells were attributed a value of 1 if (only) red squirrels had been recorded in the three years corresponding to the data used to compute the FTY, and 0 otherwise (i.e. if only grey squirrels had been recorded in the cell in those three years), and binary logistic regressions were carried out on the resulting data for 2012 and 2015. The results are summarized in Fig. 4. Both binary logistic regressions provided significant regression slope and intercept ($p < 0.001$), and there was no significant difference between regression slopes or slopes for the data corresponding to the 2012 and 2015 woodland cover data ($\chi^2 = 0.26$, $DF = 1$, $p = 0.608$ and $\chi^2 = 0.10$, $DF = 1$, $p = 0.752$ respectively). Examination of the percentiles showed that, for the 2012 data, an FTY of 1.40

(95% CI: (1.36, 1.43))) equates to a 50% chance of a red (or grey) squirrel being recorded, while, for the 2015 data, the corresponding FTY is 1.42 (95% CI: (1.39, 1.45)). Combining the data for 2012 and 2015, the binary logistic regression provided the following formulae for the probabilities of a red or grey squirrel being recorded in a cell of given FTY:

$$P(Red\ squirrel) = \frac{e^y}{1+e^y} \text{ and } P(Grey\ squirrel) = \frac{1}{1+e^y}, \text{ where } y = 4.702 FTY - 6.619.$$

For each grid cell $S_{i,j}$, the components $A_{R,Tree}^{i,j}$ and $A_{G,Tree}^{i,j}$ of the attractiveness of $S_{i,j}$ were therefore defined as

$$A_{R,Tree}^{i,j} = \begin{cases} 0 & \text{if } FTY_{i,j} = 0 \\ \frac{e^{y_{i,j}}}{1+e^{y_{i,j}}} & \text{otherwise} \end{cases} \text{ and } A_{G,Tree}^{i,j} = \begin{cases} 0 & \text{if } FTY_{i,j} = 0 \\ \frac{1}{1+e^{y_{i,j}}} & \text{otherwise} \end{cases}$$

where $FTY_{i,j}$ uses the 2015 woodland cover data, and $y_{i,j} = 4.702 FTY_{i,j} - 6.619$.

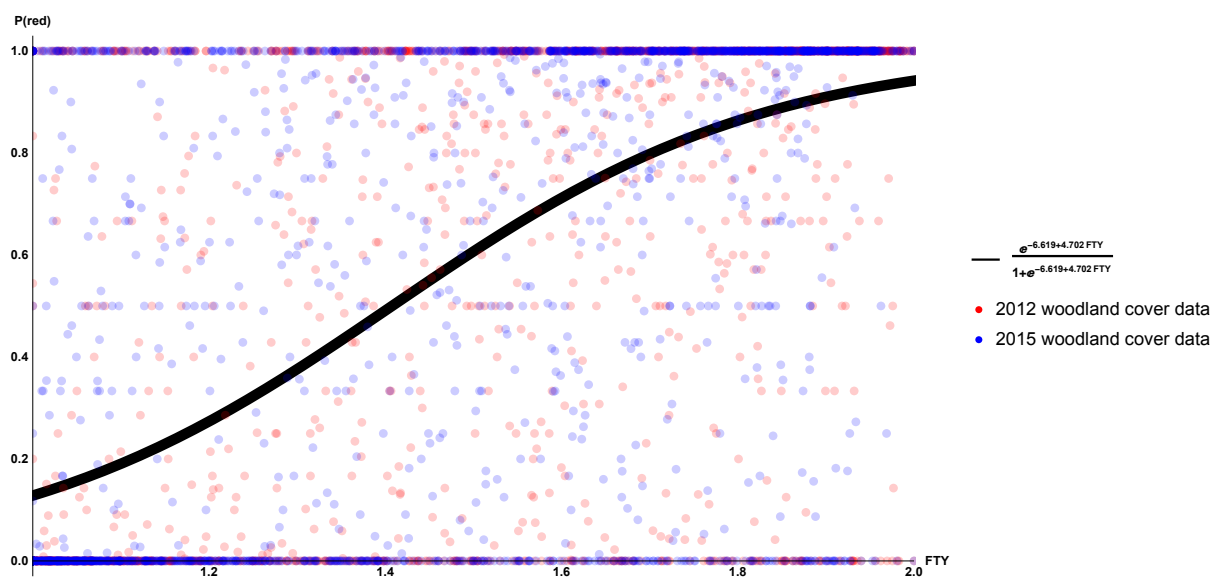

**Figure 4.** Normalized data for the squirrel records and Forest Type index (FTY) for 3,694 grid cells. Records for the years 2011 to 2013 are plotted against the FTY corresponding to the 2012 woodland cover data (in red), and records for the years 2014 to 2016 are plotted against the FTY for the 2015 data (in blue). The data is normalized to represent the proportion of red squirrels amongst all squirrel sightings reported for each grid cell. The probability of a red squirrel being reported for a cell of given FTY, derived from the binary logistic regression, is shown in black.

### 3.4. Model simulation.

Using the parameters listed in Table 1 and the values for attractiveness derived in sections 3.2. and 3.3., the model was run to simulate the evolution of the populations of red and grey squirrels in Scotland between the years 2006 and 2045. The corresponding map representations are given in Fig. 5. Representations of the reported data for selected years were also produced to test the validity of the model (see Fig. 6).

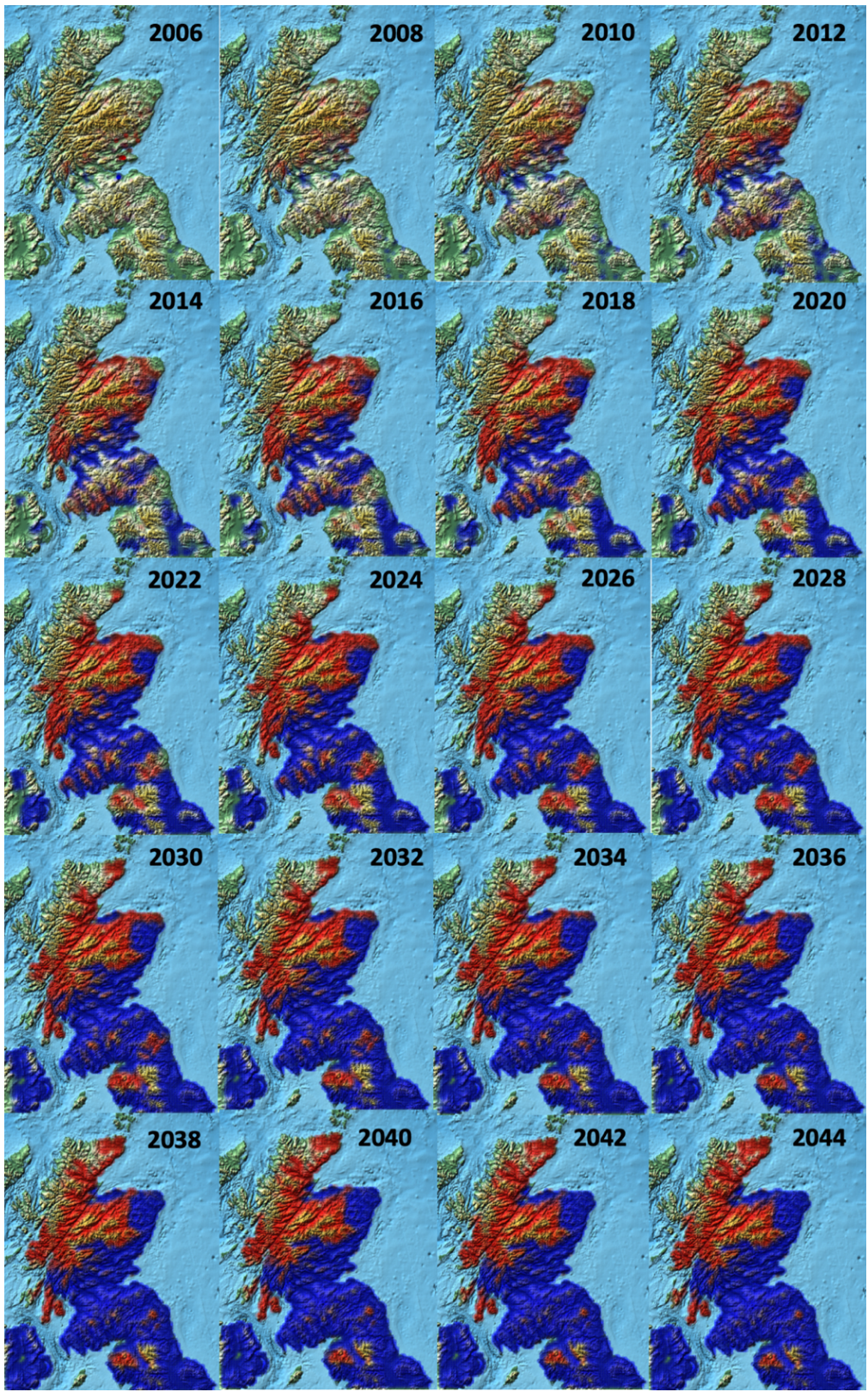

**Figure 5.** Evolutions of the red and grey squirrel populations in Scotland predicted for the years 2006 to 2044. Red squirrels are represented in red, grey squirrels in blue. Colour opacity is proportional to numbers

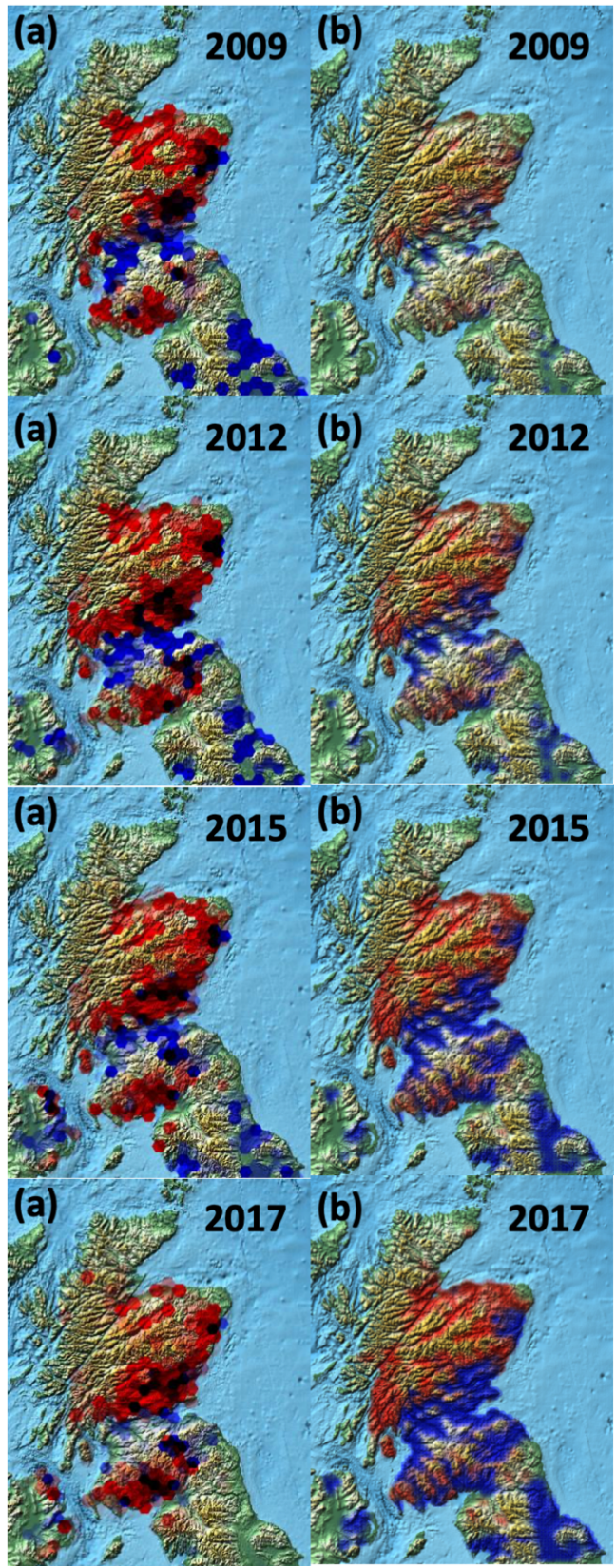

**Figure 6.** Comparison of the red and grey squirrel populations given by records and predicted by the model for the years 2009, 2012, 2015 and 2017. Red squirrels are represented in red, grey squirrels in blue. Colour opacity is proportional to numbers. (a) Number of records in Scotland. (b) Estimated population sizes in Scotland.

## *4.* Discussion:

*4.1.  Altitude and woodland cover.*

The analysis carried out showed a significantly larger number of red squirrels reported at higher altitude (over 700ft), and in forests dominated by conifers. These two phenomena might actually be linked, and reflect the fact that most conifer exploitation in Scotland occurs in the Highlands. They may also be the result of the way in which grey squirrels historically appeared in Scotland, spreading from the South and having thus not yet reached more northern area which tend to be at higher altitude and/or dominated by coniferous forests. However, they may also be indicative of different survival strategies and foraging efficiencies between the two species. Grey squirrels may not be as tolerant to low temperatures as red squirrels, which would explain how a large population of red squirrels can be maintained at higher altitude, such as in the Grampians. More generally, it has already been shown that mountain ranges, such as the Cumbrian Mountains, can provide an effective barrier to the dispersal of grey squirrels (Stevenson *et al.*, 2013). Further studies, in particular of the dens used by both species, could elucidate this issue. Regarding the composition of the wood cover, studies have demonstrated that, in areas dominated by deciduous trees, grey squirrels tend to exclude red squirrels, while both species can coexist for long periods (over 40 years) in coniferous woodland (Bryce *et al.*, 2002; Gurnell *et al.*, 2004). The positive correlation between tree cover density and FTY might simply reflect the fact that most coniferous forests in Scotland are planted and managed for timber, and tend to be much denser than either native or deciduous forests. However, such dense coniferous forests might also be better exploited by red squirrels than by grey squirrels, or might offer more refuge for red squirrels. Together, these results suggest that targeted forest management could be used in order to protect Scottish populations of red squirrels.

*4.2.  Model pertinence.*

The choice of 2006 as the year used for the initial conditions for the model was made to allow the comparison of the model's predictions with actual data until the year 2019. However, the relatively small number of records made in 2006 leads to the apparently small population sizes predicted for the first few years of the simulation (see Fig. 5). One striking consequence of the sparsity of the data available for 2006 is visible in the predictions derived for Northern Ireland. As no red squirrel sightings were reported in Northern Ireland in 2006, the model predicts the absence of red squirrels in all subsequent years, even though data collected in later years demonstrate the presence of red squirrels in Northern Ireland (see Fig. 6). Since this paper is mostly interested in the case of Scottish populations, this wasn't identified as a significant issue in this study. Furthermore, the comparison carried out between the model's predictions and data for subsequent years (see Fig. 6) shows that the model is quite realistic when it comes to Scottish squirrel populations. More accurate predictions could be obtained by including the data for further years into the initial conditions for the model, but this could have led to a form of circular reasoning and prevented any meaningful assessment of the quality of the model. Once the robustness of the model has been established, the records data for the years 2006-2019 can more safely be used to define more realistic initial conditions for the model, and thus derive further predictions. Similarly, as new data becomes available about the woodland cover, this can be incorporated into the model to refine simulations.

*4.3.   Attractiveness.*

A crucial part of the model uses the notion of attractiveness defined for each grid cell. It should be noted that this notion doesn't claim to reflect any actual biological phenomenon. For example, the fact that areas of coniferous woodland have a larger attractiveness to red squirrels than deciduous forests do should not be interpreted as an indication that red squirrels have a preference for conifers over deciduous trees. Instead, the values given for the attractiveness of each grid cell are derived from the analysis of the data available. As such, the larger attractiveness to red squirrels of coniferous woodland may only reflect the fact that red squirrels have already found refuge in coniferous forests, because they are outcompeted by the invasive greys in deciduous forests. It might also simply be an artefact of the geographical distribution of coniferous forests compared to the current limits of the grey squirrel invasion. In that sense, attractiveness may appear to be a poorly chosen name, but it remains a meaningful mathematical concept. From a modelling perspective, and irrelevant of the possible underlying biological or historical explanations, red squirrels are indeed more likely to disperse to an area covered by conifers than to a deciduous forest. The same discussion applies to differences in attractiveness based on the altitude of the grid cells.

## 5.  Conclusions

The model presented in this study is very flexible and can easily be adapted or extended. For example, it can be used for other species or different geographical areas, provided the necessary parameters can be extracted from existing data or evidence from the literature.

In the case of the interaction between red and grey squirrels, further parameters could be included when defining the attractiveness of each grid cell, such as the presence of pine marten, or the existence of local conservation measures such as trapping or culling. The interspecific competition part of the model could also be modified to include a predator-prey element to assess or simulate the effect of re-introduction of pine marten, or complemented by an SEIR model to study the impact of squirrel parapoxvirus.

Ultimately, the simulations produced should be used to better inform conservation strategies and policies in forest management, urban development and environment protection. The results obtained in this study suggest that the complete replacement of the native population of red squirrels in Scotland can be avoided, but that targeted measures might be needed, in particular the culling of grey squirrels in specific areas and the preservation and restauration of Caledonian forests.

This research did not receive any specific grant from funding agencies in the public, commercial, or not-for-profit sectors.